\documentclass[manuscript]{aastex}

\begin{document}
\pagenumbering{arabic}

\title{TEN BILLION YEARS OF GALAXY EVOLUTION}
\author{Sidney van den Bergh}
\affil{Dominion Astrophysical Observatory, Herzberg Institute of Astrophysics, National Research Council of Canada, 5071 West Saanich Road, Victoria, BC, Canada  V9E 2E7}
\email{sidney.vandenbergh@nrc.ca}

\begin{abstract}

Observations in the Hubble Deep Fields have been used
to study the evolution of galaxy morphology over time. The
majority of galaxies with $z < 1$ are seen to be disk like,
whereas most objects with $z > 2$ appear to be either chaotic
or centrally concentrated ``blobs''. Such blobs might
be the ancestral objects of ellipticals or of galaxy bulges.
About 1/3 of objects with $z > 2$ appear to be in the process of 
merging. The region with $1 < z < 2$ marks an important transition 
in the global history of star formation from a merger 
dominated regime at $z > 2$, to one at $z < 1$ in which most 
star formation takes place in galactic disks. It is speculated
that the break in the Madau plot at $z \sim 1.5$ might be related
to the transition from merger-dominated star formation at
$z > 2$ to disk-dominated star formation at at $z < 1$.
  
The Hubble tuning fork diagram only provides a satisfactory 
framework for galaxy classification at $z \leq 0.5$. At $z > 0.5$ 
barred spirals become increasingly rare. Possibly very young 
disks are still too hot (chaotic) to become unstable in global 
bar-like modes. Furthermore it becomes increasingly difficult 
to shoehorn galaxies at $z \geq 0.5$ into the Hubble classification 
sequence E - Sa - Sb - Sc. The fraction of ``peculiar'' spirals, 
i.e. those that do not fit naturally into the Hubble tuning 
fork diagram, increases from 12\% at $z \sim 0.0$ to 46\% at 
$0.6 < z < 0.8$. Early-type galaxies appear to approach their 
normal morphology faster than do late-type spirals. As a 
result only $\sim5 $\% of E - Sa - Sab galaxies are peculiar at 
$z \sim 0.7$, compared to 69\% peculiars among objects of types 
Sbc - Sc. With increasing redshift the spiral arm pattern 
in late-type spirals is seen to become ever more chaotic. In 
early-type galaxies the spiral arms appear to be less well 
developed at large redshifts than they are at $z \sim 0$. Finally 
a very a tentative, and entirely empirical, scheme is proposed 
for the classification of objects with $z > 2$.

\end{abstract} 


\section{INTRODUCTION}

As paleontologists dig down to deeper layers in the
Earth they encounter ever older fossils. By the same token
astronomers can look back in time by observing galaxies at
ever larger redshifts. The high resolution provided by the
{\it Hubble Space Telescope} (HST) makes it possible to study the
evolution of the structure within galaxies over time-scales
approaching the age of the Universe. Astronomers can now
explore the evolution of morphological classes of galaxies
in ways that are similar to those in which paleontologists 
study the evolution of fossil species and genera over time. 
Perhaps the most unexpected result of such work is that the 
Hubble tuning fork diagram only provides a satisfactory 
framework for the classification of galaxies with $z \leq 0.5$, 
i.e. for galaxies with ages \footnote {All ages used in the present paper are for a Universe in which h = 0.7,   $\Omega_m$ = 0.3 and $\Omega{_\Lambda}= 0.7$} less than half the age of the Universe. Furthermore, with
increasing look-back time, it becomes ever more difficult to 
shoehorn galaxies into the morphological sequence Sa - Sb - Sc.  

It is the purpose of the present paper to examine the
details of how the Hubble classification scheme breaks
down with increasing redshift. Part of the evolution of
galaxies over the last 10 Gyr is, no doubt, due to internal 
factors, whereas the rapidly growing frequency of mergers 
with increasing redshift is beautifully visible on HST 
images. Such growth in the merger frequency with increasing
redshift, which is expected from hierarchical merger models 
(Carlberg 2000), provides an external contribution to the 
secular changes of galaxy morphology. Strong observational 
support for a rapid increase in the merger and accretion rate 
of galaxies towards larger redshifts is provided by the 
present data, and by observations of Patton {\it et al}. (2002).

\section{CHANGES OF GALAXY MORPHOLOGY WITH TIME}

The Ohio State Bright Spiral Galaxy Survey of Eskridge {\it et al}. (2002) 
provides a more or less unbiased (van den Bergh {\it et al}. 2002) sample of nearby spiral galaxies. After degrading 101 of these images to the resolution and signal-to-noise ratio of galaxies at z = 0.7 in the Hubble Deep Field (HDF) van den Bergh {\it et al}. (2002) find that only 12\% of these objects at $z \sim 0.0$ are classified as being peculiar.  Such peculiar galaxies are, by definition, galaxies that do not fit comfortably into the Hubble classification scheme. This observed fraction of peculiar objects among nearby spirals is significantly lower than that encountered in the HDF and its Flanking Fields (FF), in which van den Bergh {\it et al}. (2000) found that 17 out of 37 (46\%) of spirals with $0.60 < z < 0.89$ are peculiar. In other words almost half of all spiral galaxies appear to be peculiar at look-back times of $\sim$ 8 Gyr. This comparison is almost independent of bandshift effects because the ground-based z = 0 images were classified in B light, whereas the classifications of galaxies at $z \sim 0.7$ were made in the infrared F814W band. In other words the galaxies at z = 0.0 and z = 0.7 were classified at almost the same rest wavelength. 
   
Van den Bergh {\it et al}. (2000) show that distant early-type
galaxies are less likely to be peculiar than is the case
for galaxies of later morphological types. Only 1/22 (5\%)
of E - Sa - Sab galaxies were observed to be peculiar at $z \sim 0.7$,
compared to 11/16 (69\%) of Sb-Sbc galaxies. These figures 
probably under-state the difference in the fraction of
peculiar early-type and late-type galaxies. This is so
because a significant fraction of ``Sc'' galaxies  are so 
peculiar at $z \sim 0.7$ that many of them have been classified
as ``proto-Sc'', ``Pec'', or ``?''. These results suggest that 
compact early-type galaxies started to approach their 
``normal'' present-day morphology faster than did more extended 
late-type spirals. However, a caveat is that it might be
more difficult to see peculiarities in distant compact early
-type galaxies than in equally distant (but more open) late-type 
spirals.

\section{BARRED SPIRALS}

Hubble (1936) noted that nearby spiral galaxies fall 
naturally into two sequences, which he dubbed ``normal spirals'' 
[Sa - Sb - Sc] and ``barred spirals'' [SBa - SBb - SBc]. One
of the first surprises (van den Bergh {\it et al}. 1996) provided 
by inspection of the {\it Hubble Space Telescope} images of the 
Hubble Deep Field North (HDFN) (Williams {\it et al}. 1996) 
was that the frequency of barred spirals appeared to decrease
with increasing redshift. This conclusion was subsequently
confirmed in the HDFN and HDFS by Abraham {\it et al}. (1999). However,
before accepting this conclusion, one has to test whether the 
apparent decrease in the frequency of bars with increasing 
redshift might be due to (1) diminishing resolution with
increasing redshift and/or (2) to a decrease in the signal-to-
noise ratio with increasing z. Furthermore such investigations
have to be carried out at constant rest wavelength to guard
against a possible wavelength dependence of the frequency
of bars (Eskridge {\it et al}. 2000). Such an investigation has
recently been undertaken by van den Bergh {\it et al}. (2002) who
compared the fraction of barred objects among a representative
sample of nearby spirals observed in blue light, with the 
fraction of galaxies classified as SB in the same sample of
galaxies, but with resolution and signal-to-noise ratio degraded
to that for HST images at $z \sim 0.7$. To compensate for  band shift 
effects these degraded images were classified in the HST 
F814W band. The data showed that $\sim$2/3 of the strongly
barred [DDO types SB0-SBa-SBb-SBc] galaxies could still be 
recognized as barred spirals at $z \sim 0.7$. In other words most
of the observed decrease in the frequency of SB galaxies
between z = 0.0 and z = 0.7 is intrinsic, and not due to
the effects of a decrease in resolution and noise with
increasing redshift. However, it should be emphasized that
such resolution and noise problems might render some bars (if
they exist) unobservable at $z \gg 0.7$.                               
   
Table 1 lists the fraction of all spirals (including
objects of type S0) with known redshifts that are barred 
in the HDF (North) from van den Bergh {\it et al}. (2000), and in
the HDF (South) from van den Bergh \& Abraham (2002). These
data show that the fraction of barred [S(B) + S(B?) + SB]
spirals decreases from $\sim$23\% for galaxies with  $z < 0.5$
to $\sim$4\% for galaxies with $z > 0.7$. As has already been noted above
all of these classifications were made at, or close to, the
ground-based B band. Taken at face value these results suggest that the frequency of barred spirals at $z > 0.5$ might be lower than it is in nearby regions of the Universe.  If confirmed by future observations, this result could be explained by assuming that the disks of spirals at large look-back times are still too chaotic (dynamically hot) to be susceptible to global bar-like instabilities. Such global dynamical instability
of disks is currently (Sellwood \& Willkinson 1993) the favored 
formation mode for bars in spiral galaxies. The notion 
that galaxies at large redshifts are more chaotic than nearby 
ones is also supported by the work of Reshetnikov {\it et al}. (2002), 
who find that disks at large redshifts are (on average) more 
often warped than are the disks of nearby galaxies. 

It is noted in passing that only $\sim$ 40\% of strongly barred 
galaxies with I(814W) $< 22$ mag, degraded to the resolution and 
signal-to-noise ratio of the HDF FF (corresponding to an 
integration time of two orbits), were recognizable as barred
spirals. This clearly shows that the low signal-to-noise  
ratio in the Flanking Fields makes it more difficult to 
recognize bar-like features in distant galaxies. 
   
In the original version of this paper it was noted that the
HDF(South) galaxy J223302.81-603325.2, at a photometric
redshift z = 1.578, is the only presently known barred spiral
at a large redshift. However, a more recent photometric redshift
of this object that is posted on the Stony Brook web site (see http:// www.ess.sunysb.edu/astro.hdfs) appears to show that the photometric redshift of this object is actually only z = 0.76. At this much lower redshift it is less surprising to find a barred spiral.

\section{GRAND-DESIGN SPIRALS}

Luminous ``grand-design'' spiral galaxies of DDO luminosity 
classes I, I-II, and II have large dimensions and are therefore
relatively easy to recognize at great distances. The combined
data for the HDF(North) by van den Bergh {\it et al}. (2000) and for
the HDF(South) by van den Bergh \& Abraham (2002) contain 
classifications for 110 spirals of known redshift, of which 
only seven are grand-design objects. None of these grand-design
spirals have $z > 0.7$. Possibly this lack of very distant grand-
design spirals is linked to the observation that the spiral
structure of late-type galaxies appears to become increasingly
chaotic as one moves to larger redshifts.  However, it is not yet
safe to conclude that grand-design spirals only exist at low 
redshifts. This is so because a Kolmogorov-Smirnov test on  
small samples described above shows no statistically significant
difference between the redshift distributions of ordinary spirals
and of grand-design spirals. So it is not yet possible to say at 
which epoch luminous disk galaxies started to form their grand-
design spiral features. It would clearly be of great interest to 
look into this question in more detail by obtaining redshifts 
(and accurate morphological classifications) for much larger 
samples of distant grand-design spirals. A possible complication 
is provided by the fact that some interacting objects at $z > 0.7$
exhibit tidal features that might be confused with the chaotic
spiral arms that are diagnostic of the most distant known
spiral galaxies.

\section{FREQUENCY OF CLASSIFICATION TYPES}

A number of authors (Brinchmann {\it et al}. 1998, Lilly {\it et al}. 1995,1998, van den Bergh 2001) have considered the possible evolution of the frequency distributions of galaxy types over time. In particular Lilly {\it et al}. (1995) and van den Bergh (2001) concluded, from observations of galaxies in the Canada-France Redshift Survey,  that there is very little change in either the density (or in the luminosity) of the reddest (early-type) galaxies over the entire range $0 < z < 1$. However, this conclusion should be treated with caution because changes in the space density [and hence galaxy morphology along any particular line of sight] may be affected by the well known correlation (Dressler 1980) between the frequency of classification types and space density. To avoid such problems in the future it will be necessary to undertake deep homogeneous redshift and classification surveys of numerous galaxies at many positions on the sky. Only in this way will it eventually be possible to gain better insight into systematic changes in the relative frequency distributions of different galaxy types with look-back time.

\section{MERGERS}

Both the HDF(North) and the HDF(South) are crowded fields. 
It is therefore expected (and spectroscopy confirms) that a 
significant fraction of all close binary galaxies are optical, 
rather than physical, pairs. Galaxies were therefore only 
classified as a ``Merger'' if (1) their main bodies overlapped or
(2) if their outer structures appeared to show evidence for 
physical overlap or tidal distortion. Since faint outer tidal 
features are probably ``lost in the noise'' in the short exposure 
Flanking Fields. Only classifications in the Hubble Deep Fields 
themselves (van den Bergh, Cohen \& Crabbe 2001, van den Bergh \& 
Abraham 2002) were used to estimate the fractions of merging 
galaxies that are listed in Table 2. These data show that about 
5\% of galaxies with $z < 1.2$ in the Hubble Deep Fields appear to 
be mergers. For comparison (see Table 2) van den Bergh \& Abraham 
(2002) find that 62 out of 109 (57\%) of their sample of HST 
galaxies with (mostly photometric) redshifts of $z > 2.0$ appear to
be mergers. In other words the typical galaxy at $z < 1$ (age $< 8$ Gyr) is single, whereas the average object at $z > 2$ (age $>10$ Gyr) appears to be merging.

\section{GALAXY MORPHOLOGY AT $z > 2$}

The majority of galaxies with $z < 1$ are observed to be disk 
galaxies. On the other hand a large fraction of the galaxies with
$z > 2$ are best described as resembling centrally concentrated ``blobs'', many of  which appear to be interacting (or merging) with other blobs. Probably these blobs are the progenitors of our present day galactic bulges and elliptical galaxies. This speculation is supported by the observation that the objects seen at $z > 2$ typically have quite small linear dimensions (Ferguson, Dickinson \& Williams 2000). Furthemore fitting of multi-color observations to evolutionary models (Popovich, Dickinson Ferguson 2001) suggests masses of $10^{9}$-$10^{11}$ $\Omega_\odot$ for the luminous inner regions of Lyman break galaxies (LBGs). Such values are diagnostic for the cores of galaxies rather than rather than of giant young star clusters. Clustering properties of LBGs have so far placed placed only weak constraints (Primack, Wechsler \& Somerville 2002) on the masses of the halos of Lyman break galaxies.

\subsection{Morphology at $z < 1$} 

As has been noted above galaxy morphology starts to deviate
noticeably from that of the templates of the Hubble classification system at redshifts $z > 0.5$. At $z \geq 1$ it becomes increasingly difficult to shoehorn most galaxies into the Hubble classification system. In the range $1 < z < 2$ galaxy morphology appears to be complex, with some characteristics that are reminiscent of the the Hubble system and others that are similar to those of the galaxies at $z > 2$.  The change in the global rate of star formation between $z \sim 1$ and $z \sim 2$ (e.g. Madau, Pozetti \& Dickinson 1998, Barkana 2002) might be due to the change in the mode of star formation from one that is merger dominated at $z > 2$ to a disk dominated phase at $z < 1$.  It is, of course, important to realize that the history of star formation may have been very different from the history of mass assembly and its subsequent re-arrangeement.

\subsection{Morphology at $z > 2$} 

The recent NICMOS $H_{160}$ images of galaxies with $2 < z < 3$ by 
Dickinson (2000) show that (after correcting for a difference in
linear resolution) the morphology of galaxies is surprisingly
similar at $\lambda_o = 1700$\AA ~and at $\lambda_o = 4300$\AA. This result suggests that the apparent structure of the images of such distant galaxy is
mainly determined by their intrinsic morphology and is not a strong function of bandshift effects. This result shows that: (1) The stars that dominate the light at $\lambda$ $\lambda_o 1200 - 1800$\AA~ also dominate at   $\lambda$ $\lambda_o 4000 - 5500$\AA. (2) If components of differing age are present they must usually have a rather well-mixed spatial distribution. (3) Dust extinction cannot be the dominant factor determining the overall structure of most galactic images at high redshift. In view of these results it appears safe to neglect band shift effects  for first-order studies of the morphology of galaxies with $z > 2$.  

Many of the galaxies at $z > 2$ are so different from those at low
redshifts that it would be quite meaningless to assign Hubble types to them. Proceeding in an entirely pragmatic fashion van den Bergh and Abraham (2002) have established an empirical classification scheme for these objects which is illustrated in Figure 1. The tentative classification system of van den Bergh \& Abraham has two parameters. The first of these indicates whether a system appears to be single (s), binary (b) or multiple (m). It should, of course, be emphasized that such a classification system is somewhat arbitrary because a member of a wide binary might reasonably also be considered a single. Furthermore a system with two bright and one faint components might be classified as a binary, rather than as a multiple system. The second classification parameter describes the structure of the brightest component of the system. Small, compact, circular knots are denoted by ``1'', slightly fuzzy brightest components are classified ``2'', fuzzy images with a comma-like structures are assigned to class ``3'', and tadpole-like objects with a longer tails are designated type ``4''. Finally chain galaxies (Cowie, Hu \& Songaila 1995) are assigned to class ``5''. It is not yet clear if the formation of such galaxy chains is in any way related to the filamentary structures that are found in numerical simulations of the early evolution of cold dark matter. It, perhaps, should be emphasized that the classification types ``1'', ``2'' and ``3'' refer to the structure of the brightest knots within structures, which may themselves be quite chaotic.

In the nomenclature adopted in the present paper ``s1'' denotes 
a compact single object,``b2'' a binary with a slightly fuzzy 
brightest component and ``m3'' a multiple object with a comma-like 
brightest component. For objects of types 4 and 5 the s, b and m 
designations are not appropriate. It is not yet clear if the strictly empirical system described above correlates with luminosity, integrated spectral type or any other physical characteristic of the objects being classified. In this connection it is, however, of interest to note that all 9 of the most distant $(z > 3.5)$ objects classified by van den Bergh \& Abraham are single objects of types s,  s1 or s2. The lack of outer ``fuzz'' in these objects might, at least 
in part, be due to the effects of cosmological $(1 + z)^4$ dimming of the faint outer regions of these ``galaxies''. From a morphological point of view the most distant objects in the HDF(North) and the HDF(South) might also be assigned to Hubble types E0 - E3 or Sa. It is of interest to note that none of the objects with $z > 3.5$ 
appear to be as flattened as edge-on spirals or E4-E7 galaxies. 
This result is somewhat puzzling because widely accepted cold dark matter evolutionary scenarios (e.g. Kauffmann, White \& Guideroni 1993, Baugh {\it et al}. 1998) form ellipticals from merging disk galaxies. Why do we not observe such disks at large look-back times? Possibly (Weil,Eke \& Efstthiou 1998) such stellar disk would be destroyed by frequent mergers at $z \gg 2$ and by stellar feedback processes that suppress cooling of gas before $z \sim 1$. The destruction of disks by tides and mergers at large look-back times might account for the unexpectedly 
small half-light radii of galaxies at $z > 1.5$ (e.g. Roche {\it et al.} 1998). 

If the speculation that the formation of disks is impeded by frequent mergers and tidal effects is correct, then one would not expect large spirals (such as the Milky Way) to contain a very old thin-disk population component.

\section{CONCLUSIONS}

The main results of investigations of the dependence of
galaxy morphology on redshift may be summarized as follows:

\begin{itemize}

\item A major transition in the dominant mode of galaxy and star 
formation is observed to occur between $z \sim 1$ and $z \sim 2$.  At low redshifts most star formation is presnetly taking place quiescently in well-ordered disks.  On the other hand much of the star formation at high redshifts appears to be associated with chaotic mergers.

\item This change might contribute to the change in the slope
of the Madau plot that is observed at $z \sim 1.5$. 

\item The majority of objects with $z < 1$ are disk-like, whereas
most of those with $z > 2$ appear to either be chaotic, or
have dominant bulge-like components. A non-exhaustive list of possible scenarious includes:  (1) the first disks were destroyed by frequent 
mergers, (2) that the gas in these disks did not cool 
before $z \sim 1$, or (3)  early gaseous disks were destroyed
by the strong winds resulting from massive starbursts. 

\item The Hubble tuning fork diagram only provides a good
framework for the classification of galaxies with $z \leq 0.5$.

\item Study of artificially degraded images shows that the
fraction of all galaxies that are barred spirals decreases with increasing redshift. This effect is tentatively interpreted as being due to the fact that
hot (chaotic) disks might not be able to develop global
bar-like instabilities.

\item The fraction of peculiar galaxies, i.e. objects that do
not fit naturally into the Hubble classification scheme
increases from 12\% at z = 0.0 to 46\% at $z \sim 0.7$.

\item Compact early-type (E-Sab) galaxies appear to approach their
final morphology more rapidly than do more extended
late-type (Sbc-Sc) galaxies. As a result only $\sim$5\% of early
type galaxies at $z\sim 0.7$ appear peculiar, compared to 69\%
peculiars among late-type galaxies.

\item At look-back times greater than 5 Gyr $(z > 0.5)$ the spiral
arm structure of early-type galaxies appears to be under-
developed.

\item At look-back times greater than 7 Gyr $(z > 0.8)$ the  
structure of the spiral arms of late-type galaxies appears less regular (more chaotic) than it does at lower redshifts.  It should be emphasized that this is an easily observed gross effect that is not likely to be affected by either noise or image resolution.

\item A tentative, and strictly empirical, classification scheme is 
proposed for those objects that are viewed at look-back times
greater than 10 Gyr $(z > 2)$. 

\end{itemize}

It is a pleasure to thank numerous colleagues who, during
the last six years, have contributed to my musings on the 
evolution of galaxy morphology over time. Particular thanks go
to Bob Abraham and Judy Cohen for their encouragement and
assistance and to Chuck Steidel for cosmological consultations
and to Simon Lilly for interesting suggestions. I am also 
indebted to David Duncan for drawing Figure 1.

\begin{deluxetable}{llll}
\tablewidth{0pt}

\tablecaption{\sc{FREQUENCY OF BARRED SPIRALS IN HDF (N) AND HDF (S) \tablenotemark{1}}}

\tablehead {\colhead{Type} & \colhead{$z < 0.50$} & \colhead{$0.50 < z < 0.70$} & \colhead{$z > 0.70$}}

\startdata

S          &    $~~~~27.5~~~~(77\%)$   &     $~~~~39~~~~(95\%)$   &   $~~~~23~~~~(96\%)$ \\
S(B)+S(B?) &    $~~~~~2~~~~~~(~6\%)$   &    $~~~~~~2~~~~(~5\%)$  &  $~~~~~~1~~~~(~~4\%)$ \\
SB         &    $~~~~~6~~~~~~(17\%)$   &    $~~~~~~0~~~~(~0\%)$  &  $~~~~~~0~~~~(~~0\%)$ \\

\enddata

\tablenotetext{1}{Includes galaxies of type S0}
\end{deluxetable}

\clearpage

\begin{deluxetable}{lcccc}
\tablewidth{0pt}

\tablecaption{\sc{DEPENDENCE OF MERGER FREQUENCY ON REDSHIFT FREQUENCY \tablenotemark{1}}}

\tablehead {& \colhead{$z < 0.5$} & \colhead{$0.5 < z < 0.7$} & \colhead{$0.7 < z$} & \colhead{$z >2$}{\tablenotemark{3}}}

\startdata

Single       &     $75$      &     $79$       &        $61$    & $41$\\
Merger \tablenotemark{2}     &     $ 4$      &     $3.5$      &        $ 4$    & $68$\\
Percent Mergers & $5\%$      &     $4\%$      &        $ 7\%$  & $62\%$\\

\tablenotetext{1}{Combined data for HDFN (excluding FF) and HDFS}
\tablenotetext{2}{Includes ``Merger?''}
\tablenotetext{3}{Objects classified as ``?'' excluded from statistics}

\enddata
\end{deluxetable}

\clearpage

\centerline{Figure Caption}

1. Illustration of a very tentative empirical classification scheme used to describe the morphologies of objects with $z > 2$.

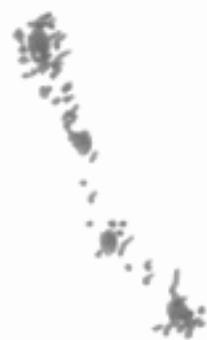
\end{document}